\newcommand{\eps}{\varepsilon}
\newcommand{\noin}{\noindent}
\newcommand{\nonu}{\nonumber}
\newcommand{\bea}{\begin{eqnarray}}
\newcommand{\eea}{\end{eqnarray}}
\newcommand{\m}[1]{{{({\bf #1})}}}
\newcommand{\au}[3]{{{\eta_{\;#1}^{(#2)}({\bf #3})}}}

\newcommand{\dde}[2]{{{\partial #1 \over \partial #2}}}

\def\bull{\vrule height .9ex width .8ex depth -.1ex }

\def\bull1{\vrule height .4ex width .4ex depth -.1ex }

\documentstyle[tighten,aps]{revtex}

\begin{document}
\draft
\title{Zero Modes and Conformal Anomaly in Liouville Vortices}
\author{G. Nardelli}
\address{Dipartimento di Fisica, Universit\`a di Trento,
38050 Povo (Trento), Italy \\ INFN, Gruppo Collegato di Trento, Italy}
\author{M. Peloso}
\address{S.I.S.S.A., via Beirut 2/4, 34014 Miramare di Trieste
(Trieste), Italy\\
INFN, Sezione di Trieste, Italy} \maketitle
\vskip 1.5truecm

\begin{abstract}
The partition function of a two dimensional Abelian gauge model reproducing
magnetic vortices  is discussed in the harmonic approximation.
Classical solutions exhibit conformal invariance, that is broken by statistical
fluctuations, apart from an exceptional case. The
corresponding ``anomaly''  has been evaluated. Zero modes of the thermal
fluctuation operator have been carefully discussed.

\end{abstract}  \vskip 3truecm
Preprint SISSA/146/99/EP
\vskip 1truecm
\pacs{11.10.Lm,11.10.Kk,11.27.+d}

\section{Introduction}

In the recent years, a great interest has been devoted to the study of magnetic
vortices, which are intimately connected to the study of
superconductivity \cite{till}. A magnetic vortex can be represented as an
infinitely long magnetic
 flux tube in
three spatial dimensions or, equivalently, as a two dimensional localised
magnetic field source. The first vortex solutions were discussed in 1957 by
Abrikosov\cite{abr} in the framework of the Ginzburg-Landau model, whereas
Nielsen and Olesen \cite{no} in 1973 were the first to recognize that the same
type of  vortices are also present in a classical relativistic model, {\it i.e.}
the Abelian Higgs model.
   Several years later Hong, Kim and Pac \cite{hkp} and,
independently, Jackiw and Weinberg \cite{jw}, discussed vortex solutions in the
same (planar) model, but with dynamics for the gauge fields governed by a
Chern--Simons (CS) term. In a CS system, the magnetic field is proportional to
the charge density, so that any excitation carrying magnetic flux is necessarily
charged, contrary to the vortices in the Abelian Higgs model, that
are electrically neutral.

Unfortunately, in both  cases the classical solutions are  known either
asymptotically or by   series  with   complicated
recursion formulas \cite{devega}. Consequently,    quantization or
thermal fluctuations  of the
 system around  the classical background are very difficult to study.

More recently, a different type of  magnetic vortex was proposed
\cite{nar} as a solution of  $2$ dimensional euclidean scalar
electrodynamics with topological coupling.  In such (Liouville) vortices, the
magnetic field satisfies the Liouville equation, all of whose regular solutions
can be easily expressed  in terms of an arbitrary analytic function.
In spite of the seeming resemblance of this model with the Abelian Higgs
system, Liouville vortices are very different from the ones in the Ginzburg
Landau theory. In the Ginzburg Landau case, the potential  of the Higgs field
leads to a spontaneously symmetry breaking. Due to this fact the classical
vortex solutions, besides the expected long range tail, have an exponentially
decreasing component, and the coefficient of the quadratic term in the Higgs
potential is related to the characteristic length of the exponential behavior.
On the contrary, in Liouville vortices the potential of the scalar field is
that of a pure $|\phi|^4$ theory, without quadratic-mass term. In addition,
there is also a non-minimal (topological) interaction that couples the matter
density directly to the magnetic field. As a consequence of these two facts,
Liouville vortices exhibit conformal invariance, and their asymptotic behavior
is always inverse power-like.

The model leading to Liouville vortices  shares
important properties with two other models: the Jackiw and Pi model \cite{jac}
and the non linear sigma model (NLSM) with $O(3)$ local symmetry
\cite{nar}. Concerning the first, the   profile of the magnetic
field  of these Liouville vortices is identical to that discussed by Jackiw and
Pi  as the static soliton solution of the gauged non linear Schroedinger
equation on the plane, in strong analogy with the  Abelian Higgs model, that is
a classical field theory whose equations of motion coincides with the non linear
Schroedinger equation governing the Ginzburg-Landau theory for type II
superconductors.  Concerning the NLSM, in ref. \cite{nar} it was shown  that all
the solutions of the above mentioned $2$ dimensional euclidean scalar
electrodynamics can be obtained from the solutions of NLSM with local symmetry.
In addition, there are other important features shared with the euclidean NLSM
in $2$ dimensions \cite{poly2} that will be discussed below.

Statistical mechanics is a natural framework for topologically non trivial
euclidean theories and, on the other hand,  many important physical
features that
the model in ref. \cite{nar} should {\it hopefully}  exhibit, are strictly
related to statistical mechanics. As an example, if these Liouville vortices are
really related to superconductivity, a phase transition of the system should
occur at some critical temperature.

In euclidean models, with positive
defined actions $S$ and  admitting  topologically non trivial solutions,   the
partition function can be defined as the path integral over the configuration
space  of the  Boltzmann factor $e^{-S}$ \cite{poly2}. In this context, the
action plays the role of  potential energy and the free energy is usually
interpreted as interaction energy  due to
``thermal'' fluctuations. However, in these cases, the definition of temperature
in not always straightforward. For instance, in the NLSM it is common to
consider the classical action already implemented by an overall factor  $1/e^2$,
{\it  i.e.}
\begin{equation}
\label{nlsm1}
S={1\over 2 e^2} \int d^2 x \partial_\mu N^a \partial_\mu N^a\ ,\qquad
N^aN^a=1\ ,
\end{equation}
and such a factor is eventually interpreted as ``thermal bath'' $\beta=1/e^2$.

As we shall see,  in  our model it will be possible to rescale the fields
in such a way that the  action  depends on the $U(1)$ coupling
$e$  only  through an overall factor $1/e^2$, just like in the NLSM above,
allowing a temperature-like interpretation of the coupling constant.

 The great
advantage of such Liouville vortices is that the classical
 solutions are given by relatively simple expressions, so that the
problem of the thermal fluctuations of
   the classical solutions can be
faced. To provide a first insight in this direction is the aim of the present
paper.

In Sect. II we shall briefly review the  model. The solutions of this
model are conformally invariant and can be classified according to their
vorticity, which is proportional to the topological invariant of the model and
is nothing but the degree of the arbitrary analytic function $\omega (z)$ upon
which the solutions depend.

In Sect. III we  begin the  study of  the thermal  fluctuations of the model.
After noticing some remarkable similarities with the euclidean  NLSM in 2D, we
consider the partition function in the so called harmonic approximation:
fields are parametrized as  classical solutions plus small thermal
fluctuations (small coupling regime), and then the action in the
Boltzmann factor is replaced by its expansion around the classical solutions up
to the second order in the thermal fluctuations.
  All the zero modes of the
corresponding fluctuation operator can be explicitly evaluated, for any
choice of
the arbitrary analytic function $\omega (z)$ characterising the  background.

If the vorticity  ${\cal N}$ of the  background is different from
the minimal one ({\it i.e. } ${\cal N}\ne 1$), the
conformal symmetry of the classical solutions is broken by thermal fluctuations,
and in Sect. IV the corresponding  ``conformal anomaly'' is explicitly
calculated.

In Section V  we discuss the special case ${\cal N}= 1$ and we comment on
future developments.

\section{The classical model}

Let us consider the 2-dimensional  euclidean action of a $|\phi|^4$ complex
scalar field minimally coupled to a Maxwell gauge potential with an additional
topological coupling
\begin{equation}
\label{naz1}
S = \int d^2{\bf x}\,{\cal L} = \int d^2 {\bf x} \left[
\frac{1}{4}F_{\mu\nu}F_{\mu\nu} + (D_\mu\phi)^*D_\mu \phi \mp e \phi^* \phi
\eps_{\mu\nu} F_{\mu\nu} + {e^2\over 2} (\phi^* \phi)^2 \right]
\end{equation}
where $\;D_\mu \phi
= \partial_\mu \phi -ieA_\mu \phi\;$ is the $U(1)$  covariant derivative and $e$
is the Abelian   coupling constant that in 2D has dimensions 1 in mass
units. The topological coupling $e \phi^*\phi \eps_{\mu\nu} F_{\mu\nu}$ is
consistent with both $U(1)$ gauge invariance and $ISO(2)$ invariance of the
action (\ref{naz1}). Actually, any general coupling of the type $e f(\phi^*\phi)
\eps_{ij} F_{ij}$ and any general scalar potential $V(\phi^*\phi)$, $f$ and
$V$ being  arbitrary functionals of the scalar density $\phi^*\phi$, would be
also allowed, due to the fact that in 2D the scalar field $\phi$ is
dimensionless. Such a general case has been considered in ref. \cite{nar} and,
with a suitable choice of the two functionals, it is possible to obtain
classical solutions of the generalised version of the  action (\ref{naz1})
satisfying a wide variety of 2 dimensional conformal equations.
Here we shall restrict to the most interesting case  given by
(\ref{naz1}).

Up to a total derivative, $S$ can be rewritten as
\begin{equation}\label{naz2}
S=\int d^2 {\bf x}\,{\cal L} = \int d^2 {\bf x}
\left[ \Big| \left( D_x \pm i\,D_y \right) \phi\,\Big|^2 + \frac{1}{2} \left( B
\mp e\,\phi^*\phi \right)^2 \right] \geq 0\ ,
\end{equation}
$B=F_{12}$ being the magnetic field. As a consequence, (\ref{naz2}) is
extremized by field configurations satisfying the following self-duality
 (anti self-duality) conditions
\bea \label{neom2}
\left( D_\mu \pm i\,\eps_{\mu \nu}\,D_\nu \right) \,\phi = 0 \ ,\nonu\\
B = \pm e\,\phi^*\phi\ .
\eea
It is easy to verify that any  configuration satisfying (\ref{neom2}) also
solves the classical Euler-Lagrange equations.
Combining together eqs. (\ref{neom2}), the scalar
density $\phi^*\phi$ satisfies the Liouville equation
\begin{equation} \label{nlio}
\triangle \mbox{ ln }(\phi^*\phi)=- 2\,e^2\,(\phi^*\phi) \ ,
\end{equation}
all of whose regular, positive definite solutions are given by
\begin{equation} \label{nB}
\phi^*\phi = \frac{4\,\left| \omega ' \right|^2}{e^2\left[
1+|\omega\,(z)\,|^{\,2} \right]^2}\ .
\end{equation}
In eq. (\ref{nB}), $\omega (z)$ is an arbitrary meromorphic function and
$\omega'(z) = d\omega/dz$. Explicit solutions of the self-duality conditions
(\ref{neom2}) in the covariant gauge $\partial_\mu A_\mu=0$ are \cite{note1}
\bea \label{nsol}
e\,A_\mu &=& -\eps_{\mu\nu}\,\partial_\nu \ln \left[1+|\,\omega\,(z)\,|^{\,2}
\right]\ , \nonu\\
e\,\phi &=& \frac{2\,\omega' (z)}{1+|\omega (z)|^2}\ .
\eea
 The above field configuration  describes a magnetic
vortex. As a matter of fact, for any choice of the arbitrary meromorphic
function $\omega (z)$ the magnetic field is always localised. The magnetic flux
$\Phi(B)$ is quantized, as the integral of the magnetic field over the whole
space is proportional to the degree of the analytic function $\omega$, {\it
i.e.} the number of solutions $z_i = z_i (\omega)$ of the equation $\omega =
\omega (z)$, each multiplied by the appropriate multeplicity $b_i$, namely
\begin{equation}
\Phi(B) = {4\pi {\cal N}\over e}\ , \qquad {\cal N}={1\over \pi}\int d^2x
{|\omega'(z)|^2\over (1+|\omega(z)|^2)^2} =\sum_i
b_i\ . \label{flux}
\end{equation}
The integer ${\cal N}$ is usually denoted as vorticity, and is the topological
invariant associated to the classical configuration. For the self-dual
configurations we are considering, ${\cal N}>0$. A parity transformation on the
solutions maps self-dual into anti self-dual configurations and the vorticity
changes sign.

In $2$ dimensions the natural flux units are  $\Phi_0=
2\pi/Q$, $Q$ being the electric charge. Consequently, if we identify $e$ with
the electric charge $Q$, the flux is an even  multiple of $\Phi_0$.
Alternatively,
eq. (\ref{flux}) gives a
magnetic flux which is an arbitrary integer ${\cal N}$ in terms of the natural
flux units $\Phi_0\;$, provided the Abelian coupling $e$ is twice the electric
charge $Q\;$,  suggesting the idea that the matter field we are considering
should be somehow related to an electron-pair condensate (Cooper's pair).

In the remaining part of this Section, we shall discuss the classical
symmetries of the solutions (\ref{nsol}), as they will play a crucial role in
the thermal fluctuations  the model (\ref{naz1}). Besides the obvious gauge and
`Poincar\'e' $ISO(2)$ symmetries,  solutions (\ref{nsol}) possess also conformal
invariance. A conformal transformation
\begin{equation} \label{ntra}
z=x+iy \rightarrow \rho(z) = \tilde{x}(x,y) + i \tilde{y}(x,y)\;,
\end{equation}
where $\rho$ is an arbitrary analytic function, connects different solutions of
the action (\ref{naz1}).
Under the action of the  conformal redefinition (\ref{ntra}), gauge and matter
fields transform as\cite{note2}
\bea
\phi(z,\bar{z}) &\rightarrow& \tilde{\phi}(z,\bar{z})= \frac{d\rho}{dz}
\phi(\rho, \bar{\rho}) \nonu\\
A_\mu({\bf r}) &\rightarrow& \tilde{A_\mu}({\bf r})= \dde{\tilde{x}^\nu}{x^\mu}
A_\nu(\tilde{{\bf r}})\label{transf}
\eea
so that the matter density $\phi^*\phi$, the magnetic field and the self-dual
derivatives transform as densities with weight $J= {\rm det} (\partial \tilde
x^\mu/\partial x^\nu)= |\rho'(z)|^2$, namely
\bea \label{ntras}
\phi^*\phi &\to& \tilde{\phi}^* \tilde{\phi}(z,\bar{z})=
J \phi^*\phi\left. \frac{ }{ } \right|_{(\rho,\bar{\rho})} \ ,\nonu\\
\left[ \left( D_\mu+i \eps_{\mu\nu} D_\nu \right) \phi \right]({\bf r})
&\to & \left[ \left( \tilde{D}_\mu+i \eps_{\mu\nu} \tilde{D}_\nu \right)
\tilde{\phi} \right]({\bf r}) = J \left[ \left( D_\mu+i \eps_{\mu\nu} D_\nu
\right) \phi \right](\tilde{{\bf r}})\ , \nonu\\
B(z,\bar{z}) &\to & \tilde{B}(z,\bar{z})= J
B\left.\frac{ }{ }\right|_{(\rho,\bar{\rho})}\ .
\eea

>From eqs. (\ref{ntras}) it follows that both  conditions (\ref{neom2}) are
preserved under a conformal transformation.
Conformal symmetry of the solutions can be easily understood in terms of the
energy momentum tensor $T_{\mu\nu}$. The independent components of $T_{\mu\nu}$
in $2$ dimensions can be chosen to be its trace $T_{\mu\mu}$ and $T^\pm =
T_{11}-T_{22} \pm 2i T_{12}\;.$ In our specific model (\ref{naz1}) such
components are given by
\begin{equation}
\label{emtensor}
T_{\mu \mu} = {1\over 2}\left( B + e\,\phi^*\,\phi \right)
\left(B - e\,\phi^*\,\phi \right)\ ,\quad T^\pm = 2 \left( D_1 \phi^* \pm i
D_2 \phi^* \right) \left( D_1 \phi \pm i D_2 \phi \right)\ ,
\end{equation}
and we see that the
self-duality conditions (\ref{neom2}), that solve the equations of motion, also
make vanishing the whole energy momentum tensor, and in particular its trace.

 Nonetheless, the action $S$ is $not$
conformally invariant. At first sight, conformal symmetry of the solutions
could seem at
odd with the fact that the action
 explicitly depends on a dimensional parameter (the Abelian coupling $e$), and
with the fact that the action is not conformally invariant. However,
 to preserve the stationary points of the action, the Lagrangian density
transforms covariantly under conformal transformations,
 \begin{equation}
\label{covariantly}
S=\int d^2{\bf x}\:{\cal L}\m{x} \to
\int d^2{\bf x}\:J\,{\cal L}\m{x}\ .
\end{equation}
 The reason of
such a nice transformation property relies in the  fact that it is always
possible to rescale the fields in such a way that the action (\ref{naz1})
does not contain {\it any} mass scale, apart from an overall multiplying
factor, just like in the NLSM (\ref{nlsm1}).  As a
matter of fact, if we rescale the   fields, collectively denoted by   $\Phi$,
as $\tilde \Phi=e\Phi$, the classical action (\ref{naz1}) written in terms of
$\tilde \Phi$ becomes  \begin{equation}
\label{scale}
S[\Phi;e]={1\over e^2} S[\tilde \Phi;1]\ ,
\end{equation}
and, classically, all the dependence on dimensional parameters in
the action (\ref{naz1}) can be ruled out.

\section{Statistical Fluctuations}

The partition function of an euclidean system admitting topologically non
trivial solutions is defined as the path integral over the configuration space
of the Boltzmann factor  $e^{-S}$ \cite{poly2}.

There are some important analogies between the model we have described in the
previous Section and the NLSM in eq. (\ref{nlsm1}) that are definitely worth
mentioning, also in view of the fact that for the NLSM
the partition function can be estimated in the  small coupling regime
\cite{poly2}.

In both the models, all the classical solutions are expressed in terms of an
arbitrary analytic function $\omega (z)$, and  the topological
invariant expressed in terms of $\omega (z)$ is $identical$ for the two models.
In addition, due to eq. (\ref{scale}), also in our model all the dependence on
the coupling constant can be factorized in an overall factor $1/e^2$,
allowing a temperature-like interpretation of the coupling constant in the
evaluation of the partition function, just like in the NLSM.

Finally, the most remarkable analogy: following \cite{poly2}, in order to
exploit the renormalization group of the system (\ref{nlsm1}), we decompose the
$N^a$ variables of the NLSM  according to
\begin{equation}
\label{nlsm2}
N^a (x) = (1 - |\phi|^2)^{1/2} N_0^a (x) + \phi^i e^a_i\ ,
\end{equation}
where $a=1,2,3$ and $i=1,2$. $N_0^a$ is the so called ``slowly varying'' vector
and $e^a_i$ are orthogonal to it, whereas $\phi^i$ represents the ``fast
fluctuations''. As a consequence of $N_0^a e^a_i=0$ and $e^a_i e^a_j =
\delta_{ij}$, we have
\begin{eqnarray}
\label{nlsm3}
\partial_\mu N_0^a &=& B^i_\mu e^a_i\nonumber\\
\partial_\mu e_i^a &=& -\epsilon^{ij} A_\mu e^a_j - B^i_\mu N_0^a\ .
\end{eqnarray}
The variables $A_\mu $ and $B^i_\mu$ have to be considered as auxiliary
variables
characterizing the slowly varying fields. Substituting eqs. (\ref{nlsm2},
\ref{nlsm3}) in eq. (\ref{nlsm1}) and selecting terms up to the second order in
$\phi^i$ one gets
\begin{equation}
\label{nlsm4}
S^{(II)}={1\over  e^2} \int d^2 x \left[(D_\mu\phi)^*D_\mu \phi - {1\over 2}
\phi^* \phi \eps_{\mu\nu} F_{\mu\nu}\right]\ ,
\end{equation}
where $\phi=(\phi^1+i\phi^2)/\sqrt{2}$ and $F_{\mu \nu}$ is the ``field
strength'' of the auxiliary variable $A_\mu$. Notice that eq. (\ref{nlsm4})
is exactly of the same type of the quadratic term in $\phi$ of the action
(\ref{naz1}); in particular, the same topological coupling is reproduced.
However, its coefficient is one half of the one in eq. (\ref{naz1}), and
therefore eq. (\ref{nlsm4}) is nothing but the $first$ term in the r.h.s. of eq.
(\ref{naz2}), up to an inessential total derivative. Consequently, extremizing
the action $S^{(II)}$ is equivalent to impose the first set of self-duality
conditions  (\ref{neom2}). The remaining term of the action in (\ref{naz2}) (and
the second set of self-duality conditions (\ref{neom2})) provides the
``dynamics'' for the field $A_\mu$ that, instead,  in eq. (\ref{nlsm4}) has to
be considered as a  background field.

Thus, a first very  rough estimate of the partition function of the model
(\ref{naz2}) could be obtained  in the following way. In the path integral, one
could integrate only over matter  fluctuations $\varphi = \phi - \phi^{cl}$
induced  by the first term in eq. (\ref{naz2}), keeping the magnetic field fixed
and equal to its classical value $B= \phi^{*\, cl}\phi^{cl}$. Then, one
immediately gets Polyakov's results and the partition function turns out to be
that of a Coulomb gas in its plasma phase, with Debye screening, and
therefore a mass gap.

On the one hand, thefact that the original system
is equivalent to a set of massive fermions is definitely a positive result,
that confirms previous conjectures \cite{nar}. On the other hand,  a physical
interpretation of this picture in terms of vortices is  still obscure and,
more important,  it is not clear to what extent the approximation of
considering only matter fluctuations is reliable.

To proceed, one should consider fluctuations of the whole set of
self-interacting fields (matter and gauge fields). Clearly, such a problem is
much more complicated and an explicit evaluation of the partition function
becomes a formidable problem. However, even without an explicit knowledge of the
partition function, something can be done and interesting results can be
obtained.

 \subsection{Harmonic approximation}

In the evaluation of the partition function of our model, the most appropriate
approximation  is the so called
harmonic approximation, where the action in the Boltzmann factor is expanded up
to the second order around the classical solutions. This approximation is
frequently used also in quantum mechanics (stationary phase approximation) where
it leads to the one loop effective potential.

Clearly, in the harminic approximation, it is assumed that trajectories
deviating
significantly from the classical solutions have a negligible weight in the path
integral. In our case, due to eq. (\ref{scale}), this assumption  is certainly
true in the small coupling regime. Thus, we shall consider the case $\beta
=1/e^2 \gg 1$.

 Let us decompose the fields ${\hat \Phi}
= \{\hat A_1,\hat A_2,\hat \phi_1,\hat\phi_2 \}$, with $\hat\phi =(1/\sqrt{2})
(\hat\phi_1 + i \hat\phi_2)$,
 as the sum of
the classical solutions
  $\;\;{
\Phi}^{cl} = \left \{ A_1^{cl}\;,\;A_2^{cl}\;, \;\phi_1^{cl}\;,\;\phi_2^{cl}
\right \}\;$ plus  ``thermal''  fluctuations $\;\;{\bf \eta} = \left
\{ a_1\;,\;a_2\;,\;\varphi_1\;,\;\varphi_2 \right \}\;$ and let us consider
the formal expansion of the action (\ref{naz1}) around the
classical solutions up to the second order in the field fluctuations. The first
two terms of this expansion vanish, due to the Euler-Lagrange equations and to
the fact that (\ref{naz1}) vanishes on the classical solutions. We have
therefore
\bea
\label{vqazapp}
S \left[ {\hat \Phi} \right] &\simeq& \frac{1}{2} \int d^2{\bf x}\;d^2{\bf y}\;
\eta_i\m{x}\;M_{ij}({\bf x},{\bf y})\;\eta_j\m{y}\;\;, \nonu\\
M_{ij}({\bf x},{\bf y}) &=& \left. \frac{\delta^2 S}{\delta
\hat \Phi_i\m{x}\,\delta \hat\Phi_j\m{y}} \right|_{cl}
\eea
Fluctuations $\eta$ are required to be sufficiently regular
and normalizable,
\begin{equation}
\label{norma}
\| {\eta} \|^2 = \int d^2{\bf x} \left[ a_\mu
a_\mu + \varphi_i \varphi_i \right] < \infty\;.
\end{equation}
Applying eq. (\ref{vqazapp}) to our model, after straightforward calculations,
we get
\bea  \label{vqM}
S \left[ {\hat \Phi} \right] &\simeq& \frac{1}{2} \int d^2{\bf x}\,d^2{\bf y}\;
\eta_i\m{x}\:\delta^2({\bf x}-{\bf y})\,{M}_{ij}\m{y}\:\eta_j\m{y}\nonu\\
&=& \frac{1}{2} \int d^2{\bf y}\;\eta_i\m{y}\:{M}_{ij}\m{y}\:\eta_j\m{y}
\nonu\\
&=& \frac{1}{2} \int d^2{\bf y}\,\left \{\;a_\mu \left[ \left( - \Delta + e^2
\left(\phi_1^2 + \phi_2^2 \right) \right) \delta_{\mu \nu} + \partial_\mu
\partial_\nu \right] a_\nu + \right. \nonu\\
&\:+\!\!\!\!& a_\mu \left[ - 2\,e\,\phi_1\,\eps_{\mu \nu} \partial_\nu - e\,
\eps_{\mu \nu} \partial_\nu \phi_1 -\,2\,e\,\partial_\mu \phi_2 +
2\,e^2 A_\mu \phi_1 \right] \varphi_1 + \nonu\\
&\:+\!\!\!\!& \varphi_1 \left[ 2\,e\,\phi_1\,\eps_{\mu \nu} \partial_\nu
- e\,\eps_{\mu \nu} \partial_\nu \phi_1 - 2\,e\, \partial_\mu \phi_2
+2\,e^2 A_{\mu}\,\phi_1 - 2\,e\,\phi_2 \partial_\mu \right] a_\mu + \nonu\\
&\:+\!\!\!\!& a_\mu \left[ - 2\,e\,\phi_2\,\eps_{\mu \nu} \partial_\nu
- e\,\eps_{\mu \nu} \partial_\nu \phi_2 + 2\,e\, \partial_\mu \phi_1 +
2\,e^2 A_\mu \phi_2 \right] \varphi_2 + \nonu\\
&\:+\!\!\!\!& \varphi_2 \left[ 2\,e\,\phi_2\,\eps_{\mu \nu} \partial_\nu
- e\,\eps_{\mu \nu} \partial_\nu \phi_2 + 2\,e\,\partial_\mu \phi_1
+ 2\,e^2 A_\mu\, \phi_2 +2\,e\phi_1 \partial_\mu \right] a_\mu + \nonu\\
&\:+\!\!\!\!& \varphi_1\,\Big[\!-\!\Delta + \frac{1}{2}\,e^2\! \left(\phi_1^2 -
\phi_2^2 \right)\!+ e^2 A^2 \Big]\,\varphi_1 + \varphi_1 \left[ e^2 \phi_1
\phi_2 - 2\,e\,A_\mu \partial_\mu \right] \varphi_2 + \nonu\\
&\:+\!\!\!\!& \varphi_2 \left[ e^2 \phi_1 \phi_2 + 2\,e\,A_\mu \partial_\mu
\right] \varphi_1 + \varphi_2\,\Big[\!-\!\Delta - \frac{1}{2}\,e^2\!\left(
\phi_1^2 - \phi_2^2 \right)\!+ e^2 A^2 \Big]\,\varphi_2 \Big\}, \nonu\\
\eea
where $A^2=A_\mu A_\mu$ and
we omitted, for brevity, the superscript $cl$ on the classical fields in the
square brackets.  In order to
factorize the $\delta^2({\bf x}-{\bf y})$ term in the first equality of
(\ref{vqM}), some integrations by parts have been performed and, in so doing,
the fluctuation operator $M_{ij}$  does not look manifestly self-adjoint.
This is
not a problem, as the normalizability condition (\ref{norma}) always permits to
write   $M_{ij}$ as in (\ref{vqM}).

Evidently, in the harmonic approximation the evaluation of the partition
function is equivalent to the calculation of the determinant of
 $M_{ij}$ and, in turn,  the solutions of the eigenvalue problem
\begin{equation}
\label{vqaut} {M}_{ij}\m{y} \; \au{j}{n}{y} = \lambda_n \;
\au{i}{n}{y}
\end{equation}
would completely solve the problem.

 Clearly, this approach is practically impossible, due to the
complicated form of the operator $M_{ij}$. Nonetheless, even without the
explicit knowledge of the determinant,  quite often it is possible to calculate
 expectation values. This is the case of the conformal anomaly, that we
shall evaluate in Sect. V with the help of the zeta-function regularization
technique.

\subsection{Zero modes of the  thermal fluctuation operator}

>From eq.  (\ref{naz2}), $S$ is manifestly positive definite and, on the
classical
solutions, (\ref{naz2}) achieves its vanishing minima. Consequently, $M_{ij}$
has no
negative eigenvalues,   and the lowest eigenvalues are the
zero modes. In our formulation, there are many operators $M_{ij}$, depending on
the choice of the classical background that, in turn, is completely
specified by the choice of the arbitrary meromorphic function $\omega (z)$. For
a given choice of $\omega $ one can associate a topological invariant ${\cal N}$
(vorticity) to the classical solutions.
The topological invariant classifies classical vortex solutions into distinct
inequivalent classes. Obviously, in each class there exist infinite analytic
functions leading to the same vorticity.

Zero modes will strongly
depend on the specific choice of the arbitrary analytic function $\omega$
characterizing the classical background. Nevertheless, the {\it number} of
normalizable zero modes of the operator $M_{ij}$ will depend only on the
homotopy
class to which the function $\omega$ belongs.

Zero modes satisfy a first order equation which is simpler than eq.
(\ref{vqaut}) with $\lambda_i=0$, and therefore is  worth mentioning.
Such an equation is a consequence of the peculiar  form (\ref{naz2}) of the
action. Due to the self-duality conditions (\ref{neom2}), the action vanishes
when evaluated on the classical background (\ref{nsol}). On the other hand,
from eq. (\ref{naz2}) the action is written as the sum of  squares, so that
the requirement that the action vanishes up to the $second$ order in the
``thermal''  expansion $\hat \Phi=\Phi^{cl} +\eta$ (zero-modes), is
equivalent to
the requirement that zero modes $\eta$ solve  the self-duality conditions
(\ref{neom2}) expanded up to the $first$ order, {\it i.e.}
\bea \label{nulcond}
\left( D_\mu^{cl} + i \eps_{\mu \nu} D_\nu^{cl} \right) {\varphi} &=& i e
\left( {a}_\mu + i \eps_{\mu \nu} {a}_\nu \right) \phi^{cl} \nonu\\
{b}&=& e \left( {\varphi}^* \phi^{cl} + \phi^{*\;cl} {\varphi}
\right)
\eea
where $\;\;\;D_\mu^{cl} = \partial_\mu - i e A_\mu^{cl}\;\;,\;\;
{b} = \eps_{\mu \nu} \partial_\mu {a}_\nu\;\;.$ It can be verified
by direct inspection that eqs. (\ref{nulcond}) are indeed equivalent to eq.
(\ref{vqaut}) with $\lambda_i=0$.
Moreover, eqs. (\ref{nulcond}) can be decoupled  and rewritten as a Schroedinger
type problem. Taking eqs. (\ref{neom2})
and (\ref{nlio})  into account, it is not difficult to check that the magnetic
field $b$ of the zero modes has to satisfy the equation
\begin{equation} \label{nulcondB}
\Delta \left( \frac{b}{B^{cl}} \right) = - 2 e {b}
\end{equation}
that, written in terms of $b/B^{cl} = \psi$, becomes the zero-energy
Schroedinger equation   for a unit mass wavefunction $\psi$ moving in the
classical potential $V=-eB^{cl}$, {\it i.e.}
 \begin{equation}
\label{schrody}
-\frac{1}{2} \Delta \psi + V \psi = 0\ , \quad V = - eB^{cl}\ .
\end{equation}
 Consequently, by finding the zero modes of our system one
gets,  as a by-product, the zero energy  solutions
  of the above Schroedinger equation.

Zero modes of the  fluctuation operator $M_{ij}$ are associated to the
continuous
symmetries of the classical solutions. Such symmetries are: translations,
rotations, gauge transformations, conformal transformations and variations of
the arbitrary parameters upon  which the classical solutions may depend.
Practically, once a gauge choice has been picked, all such symmetries
are included in conformal transformations.

It can be  shown that, starting from
a classical solution $\Phi^{cl}$, the zero mode of the operator $M_{ij}$
associated to the continuous symmetry $\Sigma$ with infinitesimal parameter
$\sigma$  is given by the variation of $\Phi^{cl}$ under the action of the
continuous symmetry $\Sigma$ , {\it i.e.} if $\Phi^{cl}\to \Phi^{cl} +
\delta_\Sigma \Phi^{cl}$, then $\eta^\Sigma=\delta_\Sigma
\Phi^{cl}/\delta\sigma$ is the zero mode of $M_{ij}$ associated to the symmetry
$\Sigma$. The proof can be easily obtained by performing a transformation
$\Sigma$ on the Euler - Lagrange equations.

The same procedure can be also  generalised  to local symmetries, like the
gauge and the conformal ones. In this case, zero modes are obtained by
performing
an infinitesimal transformation on the classical fields: the obtained
result is a
zero mode also when the function specifying the continuous transformation is
no longer infinitesimal. Several examples will be provided below.

\subsection{Explicit evaluation of zero modes}

We begin by evaluating the zero mode associated to the gauge symmetry.
By performing an (infinitesimal) gauge transformation on the classical solutions
 we get
\begin{equation} \label{nulgau}
\eta^G = \left( \, \frac{1}{e} \, \partial_\mu \, \alpha \;;\; - \, \alpha \,
\phi_2^{cl} \;;\; \alpha \, \phi_1^{cl} \, \right)
\end{equation}

\noin It can be verified that (\ref{nulgau}) is a solution of the coupled
equations (\ref{nulcond}), and therefore a zero mode of $M_{ij}$ for any
arbitrary function $\alpha\,(x)\;$, not necessarily infinitesimal.
However, the simultaneous requirement of normalizability (\ref{norma}) and
the gauge condition   $\partial_\mu A_\mu=0$ forces  $\alpha$ to be a constant.
Introducing for later convenience complex notation $a = a_1 \, + \, i \, a_2$
for the  fluctuations of the gauge potential, the only normalizable zero
mode associated to residual gauge symmetry is then
\begin{equation}
\label{nulgau2}
\eta^G = (\varphi , a)\equiv (i\alpha \phi^{cl}\ , 0)\ .
\end{equation}

We now consider the remaining symmetry transformations of the classical
solutions.
Such symmetries are all included in the conformal one, since traslations,
rotations and variations of the parameters upon which the classical solutions
may depend can be always seen as particular cases of conformal transformations.
Actually, even the zero mode (\ref{nulgau2})
  arises from a particular conformal transformation.

Since the classical solutions are completely determined in terms of the
arbitrary
analytic function $\omega (z)$ (see eq. (\ref{nsol})),
we can write a general expression for the zero mode associated to a given
symmetry $\Sigma$ in terms of  $\Delta \omega$  and $\omega$,
$\Delta \omega$ being the variation of $\omega$ under the action of
$\Sigma$. By direct calculation we obtain the following general form of
the zero modes $\eta = (\varphi , a)$:
\bea \label{genzm}
e\, \varphi&=&{2 (\Delta\omega)'\over
1+|\omega|^2}- {2\omega'
(\omega\Delta\bar\omega+\bar\omega\Delta\omega)\over(1+|\omega|^2)^2}\ ,\\
e\, a&=&2i\bar\partial \left({\omega\Delta\bar\omega+\bar\omega\Delta\omega
\over 1+|\omega|^2}\right)\ .\nonu
\eea
Now it is immediate to check that the zero mode (\ref{nulgau2}) is a particular
case of  (\ref{genzm}), with $\Delta\omega = i\,\omega\;$. In turn,
such a $\Delta \omega$ can be always obtained through a conformal
transformation,   so that    hereafter the zero mode  (\ref{genzm})  will be
classified among the conformal ones.

The modes (\ref{genzm}) are already in the Lorentz gauge ($\partial\,a\:+\:
\bar{\partial}\,\bar{a}\,=\,0\;$ in complex notation). Among them, we have to
select only the normalizable ones by imposing eq.  (\ref{norma}). In turn,
normalizability condition (\ref{norma}) can  also be written in terms of
$\omega$ and $\Delta\omega$. After straightforward calculations, one can see
that such a requirement is equivalent to the convergence condition of the two
following integrals
\bea
\label{crit}
I_1&=&\int d^2x \left| {(\Delta\omega)'\over1+|\omega|^2}\right|^2 <\infty\\
I_2&=&\int d^2x \left| {\omega
(\Delta\omega)'-\omega'\Delta\omega \over1+|\omega|^2}\right|^2<\infty\nonu
\eea

Equations (\ref{genzm}) and (\ref{crit}) define the zero modes of the
fluctuation operator associated to the continuous symmetry $\Sigma$. It
should be
noticed that (\ref{genzm})
are indeed eigenvectors of the operator $M_{ij}$ with vanishing eigenvalue even
without specifying neither the form of $\omega$, nor its variation
$\Delta\omega$. On the contrary, the normalizability criterion (\ref{crit}) will
depend on the particular choice of arbitrary function $\omega$ as well as on the
variation $\Delta\omega$ associated to the zero mode.
As a consequence, to continue, we have to provide some specific examples.

 Let us
fix the topological number of the classical solution to be ${\cal N}$. Clearly,
there are infinite functions $\omega$ with such a degree. Here,
 we shall consider two  limiting examples, in such a way that the evaluation of
zero modes in  all the other possible choices of $\omega$   will be easily
understood as intermediate between these two cases.

The first example is the totally degenerate case, where all the vorticity ${\cal
N}$ is carried by a single vortex that, for convenience, will be located at the
origin. Then,
\begin{equation}
\label{deg}
\omega (z)=\left({z\over z_0}\right)^{\cal N}
\end{equation}
where $z_0$ is a scale introduced to render $\omega$ dimensionless, as required.
The choice (\ref{deg}) corresponds to the radially symmetric classical solutions
\bea
\label{radsym}
e\,A^{cl} &=& 2\,i\,{\cal N} \left[ 1 + \left( \frac{r_0}{r}
\right)^{2\,{\cal N}}
\right]^{-\,1} \bar{z}^{-\,1} \\
e\,\phi^{cl} &=& \frac{2\,{\cal N}}{r_0^{\cal N}} \left[ 1 + \left(
\frac{r}{r_0}
\right)^{2\,{\cal N}} \right]^{-\,1} z^{{\cal N} - 1}\nonu\\
e\, B^{cl}&=& {4{\cal N}^2\over r^2}\left[\left({r\over r_0}\right)^{\cal N}+
\left({r_0\over r}\right)^{\cal N}\right]^{-2}\nonu
\eea
where $r=|z|$ and $r_0=|z_0|$.
In this totally degenerate case, all the zero modes are easily  obtained  by
performing a conformal transformations of the type $z\to z +\chi(z)$ on
the classical solutions. The corresponding variation on $\omega$ is thus
\begin{equation}
\label{domegade}
 \Delta\omega = \omega'(z) \chi(z) = {{\cal N}\over z_0}
\left({z\over z_0}\right)^{{\cal N}-1} \chi(z)\ .
\end{equation}
We have now to investigate  on the form of the function
$\chi(z)$.  Since the classical solutions (\ref{radsym}) vanish only at the
origin and at infinity, one can easily realize that the only possible
singularities of the function $\chi(z)$ can be at the origin and at infinity: a
singularity of $\chi$ in any other point $\zeta\ne 0$ would necessarily render
the corresponding zero mode singular and not normalizable in $z=\zeta\;$. On the
contrary, singularities of $\chi$ at the origin are admissible, provided that
when inserted in (\ref{genzm}) and (\ref{crit}) they cancel against  the
corresponding zeros of the classical solutions and of the integration measure.
Thus, in this totally degenerate case, it is not restrictive to consider
functions $\chi$ as pure powers of the type  $\chi_n= g_n (z/r_0)^n$, with
$g_n$ complex coefficients, and count the number of independent functions
$\chi_n$ that render the zero modes normalizable.
For a given $\chi_n(z)$ and with $\omega (z)$ of the form (\ref{deg}), the zero
modes (\ref{genzm}) take the form
\bea \label{conmo2}
e\,a &=& \frac{2\,i\,{\cal N}}{r_0^2}\,\frac{(r/r_0)^{2{\cal N}}}{1 +
(r/r_0)^{2{\cal N}}} \:\left(\frac{\bar{z}}{r_0}
\right)^{-\,1} \times \nonu\\
&& \left \{ \left( \frac{{\cal N}}{1 + (r/r_0)^{2{\cal N}}}
+ n - 1 \right) \, \bar{g}_n \,
\left( \frac{\bar{z}}{r_0} \right)^{n - 1}
+ \frac{{\cal N}}{1 + (r/r_0)^{2{\cal N}}} \, g_n \,
\left( \frac{z}{r_0} \right)^{n - 1}
\right \} \nonu\\ \nonu\\
e\,\varphi &=& \frac{2\,{\cal N}}{r_0^2}\,
\frac{1}{1 + (r/r_0)^{2{\cal N}}} \:\left(\frac{{z}}{r_0}
\right)^{{\cal N} - 1} \times \nonu\\
&& \left \{ \left( \frac{{\cal N}}{1
+ (r/r_0)^{2{\cal N}}} + n - 1 \right) \, g_n \, \left(
\frac{z}{r_0} \right)^{n - 1}
- \frac{{\cal N}(r/r_0)^{2{\cal N}}}{1 + (r/r_0)^{2{\cal N}}} \,
 \bar{g}_n \, \left(
\frac{\bar{z}}{r_0} \right)^{n - 1} \right \} \nonu\\
\eea
Considering the asymptotic behaviour of the zero modes (\ref{conmo2}) at the
origin and at infinity, one can easily see that eq. (\ref{conmo2})
define normalizable zero modes as long as $1-{\cal N}\le n\le 1$. Thus, there
are ${\cal N}+1$ admissible values of $n$. Since the coefficients $g_n$ are
complex, two linearly independent zero modes \cite{note3} correspond to each
value of $n\,$,  and eq. (\ref{conmo2}) with  $1-{\cal N}\le n\le 1$ defines
the $2{\cal N} + 2$ zero modes of the totally degenerate case.

Notice that zero modes associated to
$ISO(2)$ symmetry and to a variation of the parameter $z_0$ in (\ref{deg}) are
 included in eq. (\ref{conmo2}) with $n=0$ (traslations) and $n=1$
(rotations if $g_1$ is purely imaginary, and variation of the parameter $z_0$
if $g_1$ is purely real). In addition, from eq. (\ref{domegade}), the gauge zero
mode (\ref{nulgau2})  is  also a particular case of eq. (\ref{conmo2}) with
$n=1$.

\medskip

As a second limiting example we shall consider the totally non degenerate case,
{\it i.e.} the choice of $\omega$ that, for a given value ${\cal N}$ of
vorticity, depends on as many free parameters as possible.
Such an $\omega(z)$ is \cite{notemittag}, for instance,
\begin{equation}
\label{nondeg}
\omega(z)= \omega_0 \prod_{i=1}^{\cal N} {(z - a_i)\over (z -
b_i)} \ ,
\end{equation}
with $ a_i\ne a_j,\  b_i\ne b_j\ \ {\rm if}\ i\ne j$ and $a_i\ne b_j$ for any
$i,j$. It depends on $4{\cal N}+2$ real  parameters (the constants
$\omega_0$, $\{a_i\}$ and $\{b_i\}$ are complex) and it corresponds to a
classical ${\cal N}$-vortex configuration, where each vortex carries  the
minimum vorticity. The explicit  location of the vortices is irrelevant to our
purposes, but in general it will be a function of all the parameters
\cite{note4}.

Let us  define
\begin{equation}
\label{cala}
{\cal A} \equiv\prod_{i=1}^{\cal N}\:(z - a_i\:) \ , \qquad
{\cal B} \equiv\prod_{i=1}^{\cal N}\:(z - b_i\:) \ ,
\end{equation}
in such a way that $\omega(z)$ and its variation under a conformal
transformation $z \rightarrow z + \chi (z)\;$ can be rewritten as
\begin{equation}
\label{den}
\omega(z)=\omega_0 {{\cal A}\over{\cal B}}\ , \quad \Delta \omega=
\omega'(z)\,\chi(z) = \omega_0 {{\cal A}\over {\cal B}} \sum_{i=1}^{\cal N}
\left( {1\over z-a_i}-{1\over z-b_i} \right) \chi(z)\ .
\end{equation}

We have to investigate on the possible form of the arbitrary function $\chi(z)$
associated to the conformal
variation $\Delta \omega$ that makes the integrals (\ref{crit}) convergent.
We first notice that, if $\chi(z)$ is regular in $z=a_i$, then also
$(\Delta\omega)'$ is regular in $a_i$. Thus, integrability of $I_1$
and $I_2$ in $z=a_i$ only requires $\chi(z)$ to be  regular at $z=a_i$.  In
$z\sim b_i$,
 $(\Delta\omega)'$
 behaves like $\chi(b_i)/(z-b_i)^3$.On the other hand, the
quantity $(1+|\omega|^2)^{-2}$ goes to zero for $z\sim b_i$ like $|z-b_i|^4$.
Thus, in order to render convergent $I_1$, $\chi(z)$ must have at least simple
zeros in $z=b_i$, and therefore it has to be of the type $\chi(z)= {\cal B}\,
f(z)$, with $f$ regular in $b_i$ and $a_i$.  Finally,  integrability of $I_1$ at
infinity requires $f(z)\sim 1/z^q$ with $q\geq {\cal N} - 1$. There is only one
possibility to have such a behaviour without introducing extra singularities in
$\Delta\omega$, that is  when the poles of $f(z)$ exactly  cancel
with the zeros of $\omega'(z)$. Consequently, $f(z)$ and $\chi(z)$ have to be of
the form
\begin{equation}
\label{chinorma}
f_{\cal N}(z)= {P_{\cal N}(z)\over {\cal A}{\cal B} \sum_{i=1}^{\cal N}\left(
{1\over z-a_i}-{1\over z-b_i}
\right) }\ ,\quad \chi_{\cal N}(z) = {\cal B} f_{\cal N}(z)\ ,
\end{equation}
where $P_{\cal N}(z)$ is an arbitrary polynomial of degree ${\cal N}$.
It is not difficult to verify that for  such a choice of $\chi(z)$ also
 $I_2$ is convergent.
Thus, $\Delta_{{\cal N}}\omega = \omega'(z)\,  \chi_{\cal N}(z)$, with
$\chi_{\cal N} (z)$ defined as in (\ref{chinorma}),   is the most general
form of
conformal variations defining normalizable zero modes. Since an arbitrary
polynomial of degree ${\cal N}$ depends on ${\cal N}+1$ arbitrary complex
parameters, the number of normalizable zero modes associated to conformal
symmetry in the totally non degenerate case is $2{\cal N}+2$, just like in the
totally degenerate case.

We conclude this section by observing that all the zero  modes in the
totally non degenerate case could have also been obtained through infinitesimal
variations of the parameters defining $\omega(z)$. Under a variation of the
parameters $b_i$, the corresponding variation of $\omega$ is
$\Delta_{b_i}\omega\equiv\delta \omega /\delta b_i =
\omega(z)/(z-b_i)$. However, such a variation renders divergent $I_1$ due to
a non integrable singularity in $z=b_i$, and infinitesimal variations of $b_i$
do not define normalizable zero modes.
Under a variation of the remaining parameters $\omega_0$ and $a_i$ we have
\bea \label{varomnondeg}
\Delta_{\omega_0}\omega\equiv\delta\omega (z) /\delta\omega_0  =
\omega(z)/\omega_0\ ,\nonu\\
\Delta_{a_i}\omega \equiv\delta \omega/  \delta a_i=- \omega(z)
/(z-a_i)\ .
\eea
Substituting (\ref{varomnondeg}) in (\ref{crit}), we have that both $I_1$ and
$I_2$ are convergent, and  eqs. (\ref{varomnondeg}) define
 $2{\cal N}+2$ normalizable zero modes. Obviously, these modes are not
independent as they are precisely of the form (\ref{den}), (\ref{chinorma}).
Actually, any linear
combination of  the zero modes associated to  (\ref{varomnondeg})
can be used as a basis for the arbitrary functions $\chi_{\cal N}(z)$ in
(\ref{chinorma}).

Finally, also in this case one can easily check that the zero modes associated
to $ISO(2)$ symmetry as well as the gauge zero mode (\ref{nulgau2}) are
contained in eqs. (\ref{den}), (\ref{chinorma}) (or, equivalently,
in eqs. (\ref{varomnondeg})).

 Starting from
these two limiting examples, it is easy to extract zero modes from any other
intermediate choice of the arbitrary function $\omega(z)\;$. In all the cases,
the number of the normalizable zero modes is $2 {\cal N} + 2$ and it is thus
only a function of the topological sector associated to the classical
background.

The thermal correction to the magnetic field due to the zero modes is
given by
\begin{equation}
\label{bzm}
b=\varepsilon_{\mu\nu}\partial_\mu a_\nu= {4\over e}\partial \bar \partial
\left({\omega\Delta\bar \omega + \bar \omega \Delta \omega\over
1+|\omega|^2}\right)\ .
\end{equation}
For instance, in the radially symmetric case, eq. (\ref{bzm})
reads
\begin{equation}
\label{bzmrs}
b={8{\cal N}^2 \over e r_0^{n}r^2\left[\left({r\over r_0}\right)^{\cal N}
+\left({r_0\over r}\right)^{\cal N}\right]^2}\Re {\rm e} \left[
g_n z^{n-1} \left( n -{\cal N}-1 +{2{\cal N}\over 1+
\left({r\over r_0}\right)^{2{\cal N}}}\right)\right]\ ,
\end{equation}
with $1-{\cal N}\le n\le 1$. It can be shown that the integral over the whole
plane of eq. (\ref{bzmrs})  vanishes for $any$ $1-{\cal N}\le n\le 1$, so that
zero modes corrections to the magnetic field do not modify the vorticity of the
classical solutions, as expected on general ground.

Having the zero modes corrections to the magnetic field, eq. (\ref{bzm}) or
(\ref{bzmrs}),  the solutions of the zero energy sector of the
Schroedinger equation (\ref{schrody}) straightforwardly follow.

\section{Conformal anomaly}

Although the overall factor $1/e^2$ in eq. (\ref{scale}) is clearly
irrelevant at
the classical level, one expects that statistically such a term is no longer
inessential, as it introduces an explicit scale in the path integral. In this
case the coupling constant will acquire a non trivial dependence on the
arbitrary mass scale $\mu$ that any regularization precedure entails
\cite{noteregu},  and an ``anomaly'' is expected: the expectation value of the
trace of the energy momentum is likely to be non-zero.
Here we shall evaluate this expectation value by using the $\zeta$-function
regularization technique.

The determinant of the thermal fluctuation operator $M$, defined as the product
of its non vanishing eigenvalues\cite{note7} ${\rm Det}\, M =\prod_n' \lambda_n
[M]$, is divergent due to the unbounded nature of its eigenvalues
$\lambda_n$.
A popular way to circumvent this problem, is to define the regularised
determinant through the zeta function. The zeta function appropriate to our
problem is defined by
$$\zeta\,\left( s | M \right) \equiv
\sum_n\,'\,\lambda_n^{-s} \left[ M \right] = \mbox{ Tr } \left( M^{-s}
\right)\ .$$
The
$\zeta$ function is regular for ${\rm Re}\, s>d/m$, where $m$ is the order of
the differential operator $M$ and $d$ the dimension of the manifold.
However, $\zeta$ can be analytically continued to the point $s=0$ and  the
regularised determinant is  defined as
\begin{equation}
\label{zdet} \left[\,\mbox{ln
Det}\,'\,M\,\right]_{\,\zeta} \equiv \lim_{s \rightarrow 0} \left[\,-
\frac{d}{d\,s}\,\zeta\,\left( s | M \right) \right] \equiv - \zeta\,' \,\left( 0
| M \right)\ . \end{equation}
Let us introduce the kernel
\begin{equation} \label{zg}
\zeta\,\left(\,t;{\bf x},{\bf y}\,|\,M\,\right) \equiv \frac{1}{\Gamma\,(s)}\:
\int_0^\infty dt\,t^{\,s -1}\,\left[\,h\,\left(\,t;{\bf x},{\bf y}\,|\,M\,
\right) - P^{\,(0)}\,({\bf x}\,,\,{\bf y})\,\right]\;\;\;,
\end{equation}
where $P^{(0)}$ is the projector into the zero mode space and
$h\,\left(\,t;{\bf x},{\bf y}\,|\,M\,\right)$ the heat kernel satisfying the
differential equation $(\partial/\partial t + M_x) h\,\left(\,t;{\bf x},{\bf
y}\,|\,M\,\right)=0$ with boundary condition $h\,\left(\,0;{\bf x},{\bf
y}\,|\,M\,\right)=I \delta^{(d)} ({\bf x}-{\bf y})$. The heat kernel at ${\bf
x}={\bf y}$ can be expanded in powers of $t$ through the so called Seeley  - de
Witt \cite{sd} coefficients
\begin{equation} \label{zsee}
h\,\left(\,t;{\bf x},{\bf x}\,|\,M\,\right) =
\frac{1}{\left( 4\,\pi\,t \right)^{d/m}}\:\left[\,a_0\,(\,x\,|\,M\,) +
a_1\,(\,x\,|\,M\,)\:t + \dots \,\right]\ .
\end{equation}
Finally,  standard manipulations
\cite{abd} permits to express the value of $\zeta (s=0 \, |\, M)$ in terms of
the Seeley - de Witt coefficient $a_{d/m}$ and the number $N$ of zero modes of
the operator $M$ as
\begin{equation}
\label{zetazero}
 \zeta (s=0 \, |\, M)= {1\over (4\pi)^{d/m}}{\rm Tr}\int d^d{\bf x} a_{d/m}
({\bf x}) - N\ .
\end{equation}
In our case, $d/m=1$ and the total number of zero modes is $N=2{\cal N}+2$, as
seen in the previous section.

It can be shown \cite{bd} that the integral of  the expectation
value of the trace of the energy momentum tensor is given by
\bea
\label{vactovac}
\int d^2 {\bf x} \left< T_{\rho\rho} \right> &=&- {\delta {\cal W}\over
\delta\log\mu}\\
{\cal W}& =& - \frac{1}{2} \lim_{s \to 0} \mbox{ Tr}\,'\,
\left[\, \frac{d}{ds} \left( \frac{M}{\mu^2} \right)^s\,\right] = \frac{1}{2}
\left[ \zeta'(0) + \mbox{ ln } \mu^2\;\zeta(0) \right]\nonu
\eea
where $\mu$ is the usual  mass parameter introduced to render the operator
$(M/\mu^2)$ dimensionless, as required. Consequently, the integrated
trace anomaly is just given by $\zeta (0)$, up to a sign,
\begin{equation} \label{acres}
\int d^2{\bf z}\;\left< T_{\rho\rho} \right> = -\,\zeta\,(0)\ .
\end{equation}
 In turn, from eq.
(\ref{zetazero}), $\zeta (0)$ can be expressed in terms of the total number of
zero modes of the operator $M$ and in term of the integrated $a_1$
Seeley - de Witt coefficient. Thus, we only need to evaluate such a coefficient.

If a (matrix valued) differential operator $M$ is  written in the form
\begin{equation} \label{a1m}
M = D_\mu^+\,D_\mu + X\;\;,\;\;X = X^+
\end{equation}
with $\; D_\mu = {\bf 1} \partial_\mu - i C_\mu$,
$\; D_\mu^+ = -\, {\bf 1} \partial_\mu + i C_\mu^+$,
$\; C_\mu=C_\mu^+\;$,
 $C_\mu$ and $X$ being matrices whose entries are solely classical
fields (not operators), then  the coefficient $a_1$ is just $-X$.
In our case the operator $M$ was already introduced in eq. (\ref{vqM}), and
after algebraic manipulations and taking the gauge $\partial_\mu A_\mu =0$
and the self-duality conditions into account, it follows that $M$ can be indeed
written in the form (\ref{a1m}), with

\begin{equation}
C_1 = i\,e \left(
\begin{array}{cccc}
0 & 0 & 0 & 0 \\
0 & 0 & -\,\phi_1 & -\,\phi_2 \\
0 & \phi_1 & 0 & A_1 \\
0 & \phi_2 & -\,A_1 & 0
\end{array} \right)
\end{equation}

\begin{equation}
C_2 = i\,e \left(
\begin{array}{cccc}
0 & 0 & \phi_1 & \phi_2 \\
0 & 0 & 0 & 0\\
-\,\phi_1 & 0 & 0 & A_2 \\
-\,\phi_2 & 0 & -\,A_2 & 0
\end{array} \right)
\end{equation}

\begin{equation}
X = \left(
\begin{array}{cccc}
0 & 0 & L_{11} & L_{12} \\
0 & 0 & L_{21}  & L_{22}  \\
L_{11} & L_{21} &
-\,\frac{3}{2}\,e^2\,\phi_1^2 -\,\frac{1}{2}\,e^2\,\phi_2^2 &
- e^2\,\phi_1\,\phi_2 \\
L_{12} & L_{22} &
- e^2\,\phi_1\,\phi_2 &
-\,\frac{1}{2}\,e^2\,\phi_1^2 -\,\frac{3}{2}\,e^2\,\phi_2^2
\end{array} \right)\ ,
\end{equation}

where
\begin{equation}
L_{\mu a}=-e \varepsilon_{ab}D_\mu\phi_b\ ,
\end{equation}
$D_\mu\phi_a=\partial_\mu\phi_a +e A_\mu \varepsilon_{ab}\phi_b$ being the
$U(1)$ covariant derivative of the matter fields $\phi_a$.

Consequently,  the trace of $a_1$ is
$$\mbox{tr } a_1\m{z}= -\mbox{tr } X = 2\,e^2\,\left( \phi_1^2 + \phi_2^2
\right) = 4\,e\,B\m{z}$$
so that its integral is a topological invariant. From
eqs. (\ref{acres}), (\ref{zetazero}) it follows that
\begin{equation} \label{acres2}
\int d^2{\bf z} \left< T_{\mu\mu}\m{z} \right> = 2 (1 - {\cal N})
\end{equation}
manifesting a trace anomaly for any ${\cal N}\ne 1$.

\section{Discussion}

Among the local gauge theories admitting  classical vortex solutions,
the one we have presented is particularly interesting. First of all its
  classical solutions are very simple. All the solutions can be expressed in
terms of an arbitrary analytic function, which is the arbitrary analytic
function specifying the regular solutions of the Liouville equation.
To any classical field configuration, there is an associated  topological
invariant (vorticity), which is the degree of the analytic function.
The topological invariant is  the same of that characterizing the solutions
of the 2D euclidean NLSM, all of whose solutions can be also expressed in terms
of an arbitrary analytic function.

The trace of the energy momentum tensor vanishes on the classical solutions
and, consequently,
 conformal transformations are a symmetry of
the classical  solutions: a conformal reparametrization relates two different
solutions of the model. Related to this fact there is also the factorization
property (\ref{scale}) of the coupling constant $e$, that is the only
dimensional parameter of the model. Such a factorization is important also
because it permits, for topological models, a temperature-like interpretation of
the coupling constant  ($\beta = 1/e^2$). In this context, the euclidean
action should be interpreted as the potential energy of the system whereas the
free energy should represent the energy fluctuations due to the interaction
with the thermal bath $\beta$. In the limit of small coupling (large $\beta$),
the partition function of the system is just the determinant of operator
obtained byexpanding the classical action around the
classical solutions up to the second order in the thermal fluctuations.
The evaluation of the determinant requires a regularization procedure and,
therefore, an arbitrary scale.  Consequently, one expects that thermal
fluctuations destroys scale (and conformal) symmetry, and in fact there is a
conformal ``anomaly'': the expectation value of the trace of the energy momentum
tensor  does not
vanish, except for the special value ${\cal N}=1$ (see eq. (\ref{acres2})).
When the classical background carries the minimum vorticity (${\cal N}=1$),
 conformal symmetry seems to survive,  at least in the harmonic approximation.

This is a very surprising property that certainly deserves a deeper analysis. At
present, we do not know the exact reason of such a phenomenon. Nonetheless,
 there are indeed some properties that make the case ${\cal N}=1$
different from the others.
For example, the trace of the $a_1$ Seeley -- De Witt coefficient or,
alternatively, the shape of the potential $V=-e B^{cl}$ felt by the Schroedinger
particle (\ref{schrody}), dramatically changes in the cases
${\cal N}=1$ and ${\cal N}\ne 1$. Let us consider for convenience the radially
symmetric case. Then, if ${\cal N}=1$, the potential $V=-eB^{cl}=-e^2
\phi^{*cl}\phi^{cl}$ is a monotonic radial function, and therefore it has a
single, non degenerate minimum at $r=0$. On the contrary, if ${\cal N}\ne
1$, the
potential  $V$ has a maximum point at $r=0$, whereas its minima are degenerate
and located around the circle  $r=r_0 [({\cal N}-1)/({\cal N}+1)]^{1/2{\cal
N}}$. Thus, there is the  intriguing possibility that the occurence of conformal
anomaly could be related to the vacuum degeneracy of the free energy.

Another feature that makes the case ${\cal N}=1$ different from all the others
is the following: the most general solution carrying vorticity $1$ is obtained
by choosing the arbitrary analytic function $\omega (z) = \omega_0
(z-a)/(z-b)$. It is easily recognized that such a function is in a one to one
correspondence with an arbitrary  transformation  of the group $SL(2,{\bf C})$,
$z\to \zeta = (a z + b)/(c z + d)$,  $ad-bc=1$. This is a
  very special subgroup of the conformal group: it defines the
projective transformations, that are the only conformal transformations
providing invertible mappings of the whole complex plane onto itself. As a
consequence of this fact, $SL(2,{\bf C})$ is a kind of ``residual'' symmetry
one has, once the the number of vortices  and the total vorticity  has been
fixed: let the pair  $(Q,{\cal N})$ denote an arbitrary background field
configuration of $Q$ vortices with total vorticity ${\cal N}$; then, applying
an arbitrary $SL(2,{\bf C})$ transformation, the background $(Q,{\cal N})$ is
mapped into another background but with the same pair $(Q,{\cal N})$.
 By contrast,   if a conformal
transformation $z\to \zeta (z)$  $not$ belonging to $SL(2,{\bf C})$ is
applied to
a field configuration  $(Q,{\cal N})$, it necessarily increases vorticity,
{\it i.e.} $(Q,{\cal N})\to (Q'\ge Q,{\cal N}'>{\cal N})$.
Clearly, if ${\cal N}=1$ (minimum vorticity) then necessarily $Q=1$, and only
in this case
 it happens that  the most general element
$\omega (z)$ characterizing the background  $(Q=1,{\cal N}=1)$ belongs to the
same  group of transformations leaving the background $(Q=1,{\cal N}=1)$
unchanged. This observation is at the root to understand why the ${\cal
N}=1$ case is not anomalous.

Besides the conformal anomaly, we have also evaluated the zero modes of the
thermal  fluctuation operator.
For a given fixed vorticity ${\cal N}$, we have derived the zero modes in two
limiting cases of the classical background: the totally degenerate and the
totally non degenerate cases, corresponding to a single vortex located at the
origin and  carrying  vorticity  ${\cal N}$, and to ${\cal N}$ distinct
vortices carrying each the minimum vorticity, respectively. All the other
possible cases can be  easily derived as intermediate between these two.
Clearly, the explicit form of the zero modes dependson the classical
background. However, the total number of
normalizable zero modes is only a function of the vorticity of the classical
solutions. In addition, the contribution to the
thermal fluctuations given by the zero modes does not change the vorticity of
the classical solution.

There are several aspects related to this model that deserve consideration for
future investigations.
An  important problem is certainly a deeper understanding
of the persistence of conformal symmetry in the ${\cal N}=1$ case, and its
possible relation with the non-degenerateness of the free energy vacuum.

Concerning the eigenvalue problem (\ref{vqaut}) or, equivalently, the
 determinant of the operator $M$,  an exact evaluation
  seems very difficult.
However, some
approximate method should be available to investigate the  system beyond
the zero-mode sector, perhaps reducing the number of degrees of freedom by
introducing some collective coordinates. Alternatively, one could try
to investigate some particular limits that simplify the form of the matrix $M$.
To this purpose, two particular cases should be mentioned:
if $r_0\to 0$, Liouville vortices becomes Aharonov-Bohm vortices, {\it i.e.}
${\bf A}^{cl}\sim \nabla \theta$ and $B^{cl}\sim \delta ({\bf r})$. This limit
greatly simplify the form of the  fluctuation operator, although in this
case the classical solutions become singular. Another interesting limit is the
large ${\cal N}$ limit; in this case the  magnetic field has a
significant non vanishing contribution only in a neighbourhood of $r=r_0$, and
it could be easier to evaluate the determinant.

Another interesting issue is the possibility that  the model  (\ref{naz1})
could be  an `effective action'  of another, more elementary, model.
This hypothesis could be supported by the fact that Liouville vortices always
have an even vorticity in terms of the elementary flux quanta
$\Phi_0=2\pi/e$ when the Abelian coupling  $e$  is interpreted as electric
constant. Consequently, it could be that the scalar field $\phi$ is related to
an electron pair condensate.  A further  point towards this direction is the
one discussed at beginning of Section 3: in the very drastic approximation
considered there, the partition function is the one of a Coulomb gas in its
Debye phase that, in turn, can be equally described by a system of massive
fermions.

Finally, at the purely classical level, it could be interesting to
investigate on
the  possibility of constructing non linear superpositions of Liouville
vortices:  just like the 't Hooft Polyakov monopole can be seen as a non linear
superposition of infinite instantons equally separated in time \cite{ros},
a non linear superposition of Liouville vortices could provide
new  soliton solutions of some lower dimensional gauge theory.


\begin{references}
\bibitem{note2}{To understand the transformation properties of the fields under
(\ref{ntra}), one can consider the solutions  (\ref{nsol}) with $\omega\equiv
z$, and then interpret the general form (\ref{nsol})  as the conformally
transformed fields under the redefinition $z \to \omega (z)$. Then, eqs.
(\ref{transf}) follow.}
\bibitem{note3}{In spite of the fact that zero modes are defined up to an
overall constant, $g_n $ and $i g_n$ define two distinct zero
modes due to the fact that in eq. (\ref{genzm}) zero modes depend linearly on
$\Delta\omega$ and its complex conjugate.}
\bibitem{notemittag}{The choice of $\omega$ is not unique. For instance, an
alternative choice is a finite Mittag-Leffler sum of the type
$\omega(z)=\omega_0 + \sum_{n=1}^{\cal N} a_i/(z-b_i)$, depending on $2{\cal
N}+1$ complex parameters. We choose the finite product expansion because in
this case is much easier to check the normalizability conditions of the zero
modes. It is important to remark that, whatever is the choice of $\omega$ when
the vorticity is fixed, the maximum number of real  parameters upon which
$\omega $ depends is  fixed and equal to $4{\cal N}+2$.}
\bibitem{note4}{The position of the vortices can be easily found by studying
the extrema of the magnetic field $eB=4|\omega'(z)|^2/(1+|\omega(z)|^2)^2$; for
example, in the ${\cal N}=1$ case the position of the single vortex is at
$z_1=(a_1 |\omega_0|^2  + b_1)/(1 +|\omega_0|^2)$.}
\bibitem{noteregu}{In this case, a regularization in the evaluation of the
partition function is required due to the infinite number of degrees of freedom
of the system.}
\bibitem{note7}{Following the standard notation, the prime index means that the
vanishing eigenvalues have to be omitted from  sums and/or products.}
\end{references}
\end{document}